\documentclass[final,5p,times,twocolumn,authoryear]{elsarticle}
\usepackage{amssymb}
\usepackage{lipsum}

\usepackage{amsmath} 
\usepackage{multirow}
\usepackage{booktabs} 
\usepackage{balance}

\journal{arXiv}

\begin{document}

\begin{frontmatter}

\title{Dosimetric Impact of Hidden Input Parameters in Inverse Optimization Algorithms for GYN HDR Brachytherapy}

\author[first]{YeongHyeon Park}
\author[first]{Shiqin Su}
\author[first]{Sarath Vijayan}
\author[first]{Zhiqian Henry Yu}
\author[first]{Mandy Cunningham}
\author[first,cor1]{Yusung Kim}

\affiliation[first]{organization={Department of Radiation Physics, The University of Texas MD Anderson Cancer Center},%Department and Organization
            state={Texas},
            country={United States}}

\cortext[cor1]{Corresponding author: Yusung Kim (ykim21@mdanderson.org)}

\begin{abstract}
\noindent \textbf{Background:} Inverse optimization (IO) algorithms are used in GYN HDR brachytherapy planning, with user parameter settings embedded in commercial TPS.
\textbf{Purpose:} To examine the dosimetric influence of hidden input parameters in three IO algorithms—IPSA, HIPO, and MCO—for GYN HDR brachytherapy across two applicator types.
\textbf{Methods:} In-house implementations of IPSA, HIPO, and MCO were implemented and evaluated against retrospectively generated commercial TPS plans (Oncentra Brachy) using identical clinical input parameters across 24 cervical cancer cases (18 T\&O; 6 T\&O+Needles (T\&O+N)). Each IO algorithm was assessed using 1,000 combinations of hidden parameters (e.g., dwell-time modulation constraints, convergence thresholds). Cumulative DVH curves and dosimetric indices (HR-CTV D98/D90, OAR D2cc) were compared with commercial plans. Standard deviations (SD) of DVH differences were used to characterize sensitivity to hidden parameters.
\textbf{Results:} For HR-CTV, SD values in T\&O+N cases reached 23.0 Gy and 7.1 Gy for MCO and HIPO, respectively, with corresponding average values of 55.8 Gy and 19.7 Gy. In T\&O cases, HR-CTV SD values reached 4.9 Gy and 3.3 Gy for HIPO and IPSA, respectively, with average values of 20.1 Gy and 8.6 Gy. MCO exhibited the highest sensitivity, followed by HIPO and IPSA. T\&O+N cases showed greater sensitivity than T\&O cases. Absolute differences in HR-CTV D90 (D98) relative to commercial algorithms reached up to 33.3 Gy (28.4) for T\&O+N cases and 10.8 Gy (8.5) for T\&O cases. For OARs, absolute D2cc differences in T\&O+N (T\&O) cases reached up to 8.6 Gy (2.3) for rectum, 17 Gy (10.2) for bladder, 14.8 Gy (3.9) for sigmoid, and 7.0 Gy (8.1) for bowel.
\textbf{Conclusions:} Hidden input parameter settings significantly impact on GYN HDR plans, with target coverage up to 28.4 Gy across IO algorithms for both T\&O and T\&O+N cases. The findings in this study shown the potential to improve plans through hidden input parameter optimization.
\end{abstract}


\end{frontmatter}

\begin{figure*}[t]
    \centering
    \includegraphics[width=0.9\linewidth]{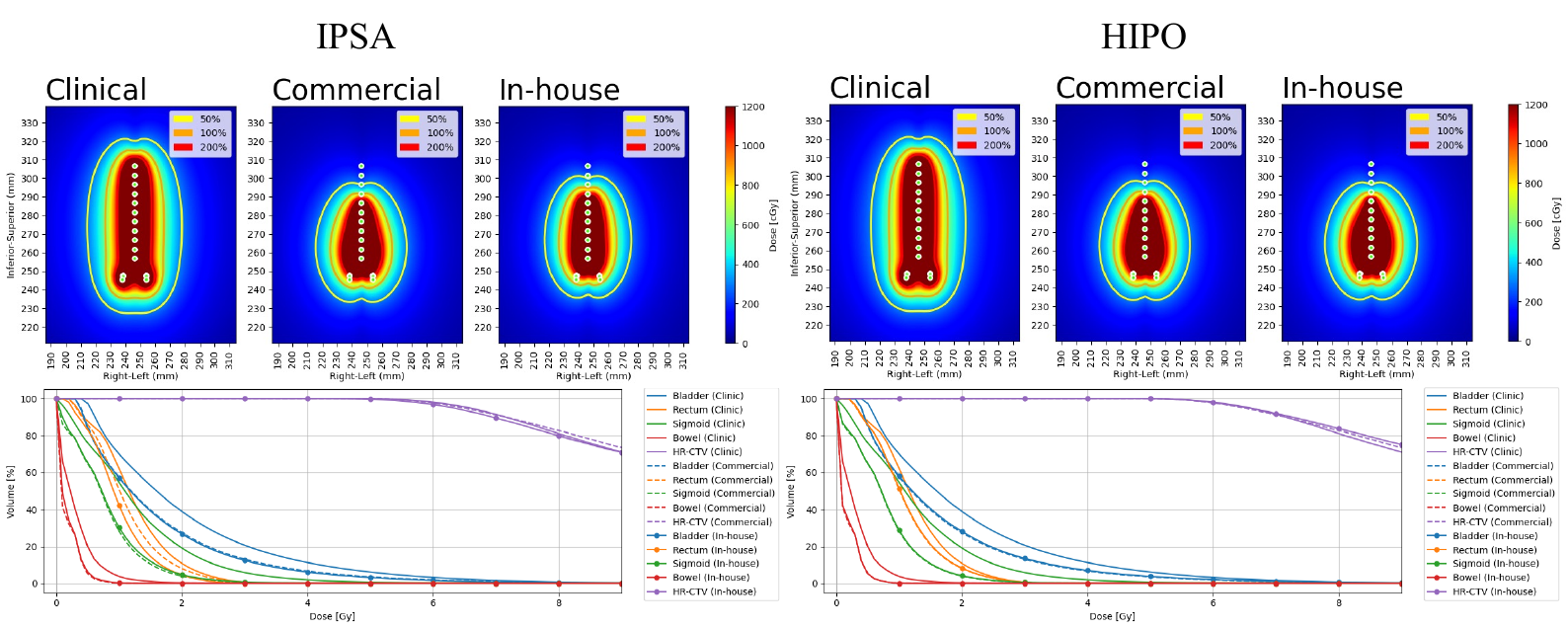}
    \vspace{-0.4cm}
    \caption{Plan results in a representative T\&O case. Dose distributions, isodose lines, and cumulative DVH curves for the clinical plan and two retrospectively generated plans with identical clinical input parameters: the commercial IO and the in-house implementation plans. Green dots indicate dwell positions.}
    \label{fig:plan_results}
\end{figure*}

\section{Introduction}
Most studies comparing inverse optimization (IO) algorithms focus on plan quality using user-defined input parameters. However, these algorithms are typically treated as ``black boxes,'' with internal parameters hidden in commercial treatment planning systems (TPS). This study explicitly assesses the dosimetric influence of hidden input parameters across inverse planning simulated annealing (IPSA)~\citep{Lessard2001,Lessard2002}, hybrid inverse planning optimization (HIPO)~\citep{Lahanas2003,Karabis2005}, and multi-criteria optimization (MCO)~\citep{Cui2018,Belanger2019} in GYN HDR brachytherapy.

IPSA was introduced as an IO approach using a simulated annealing-based stochastic optimization framework~\citep{Lessard2001,Lessard2002}. HIPO was developed as a hybrid optimization algorithm for HDR brachytherapy planning~\citep{Lahanas2003,Karabis2005}. MCO approaches have been investigated for HDR brachytherapy, including Pareto surface approximation and accelerated implementations~\citep{Cui2018,Belanger2019}.

In clinical practice, commercial TPS are commonly used, where planners adjust user-defined parameters to meet target coverage and organ-at-risk (OAR) sparing goals. Multiple studies have analyzed planning time, plan quality, or parameter settings under such clinical workflows~\citep{Jamema2011,Oud2020,Tomihara2025}. Automated or learning-based planning approaches have also been investigated for HDR brachytherapy~\citep{Shen2019,Pu2022,Rossi2025}. Commercial implementations additionally include internal configurations that are not exposed as user-defined inputs (e.g., dwell-time modulation constraints, convergence thresholds, and update regularization settings).

This study implements three inverse optimization algorithms\textemdash{}IPSA, HIPO, and MCO\textemdash{}in-house and retrospectively compares them with commercial TPS plans for GYN HDR brachytherapy. Hidden input parameters were systematically varied to evaluate the dosimetric sensitivity of DVH metrics across 24 cervical cancer cases.

\begin{table}[t]
    \caption{User-defined input parameters for commercial TPS plan generation. The same input configuration was applied to all cases and to the corresponding in-house plan generation.}
    \vspace{0.15cm}
    \label{tab:input_params}
    \centering
    \scriptsize
    \begin{tabular}{l|l|r}
        \toprule
            Structure & Dose objective (Gy) & Relative weight \\
        \midrule
            HR-CTV   & Minimum 6     & 100 \\
            Bladder  & Maximum 5     & 20  \\
            Bowel    & Maximum 3     & 20  \\
            Rectum   & Maximum 3     & 20  \\
            Sigmoid  & Maximum 3     & 20  \\
        \bottomrule
    \end{tabular}
\end{table}

\begin{figure}[t]
    \includegraphics[width=\columnwidth]{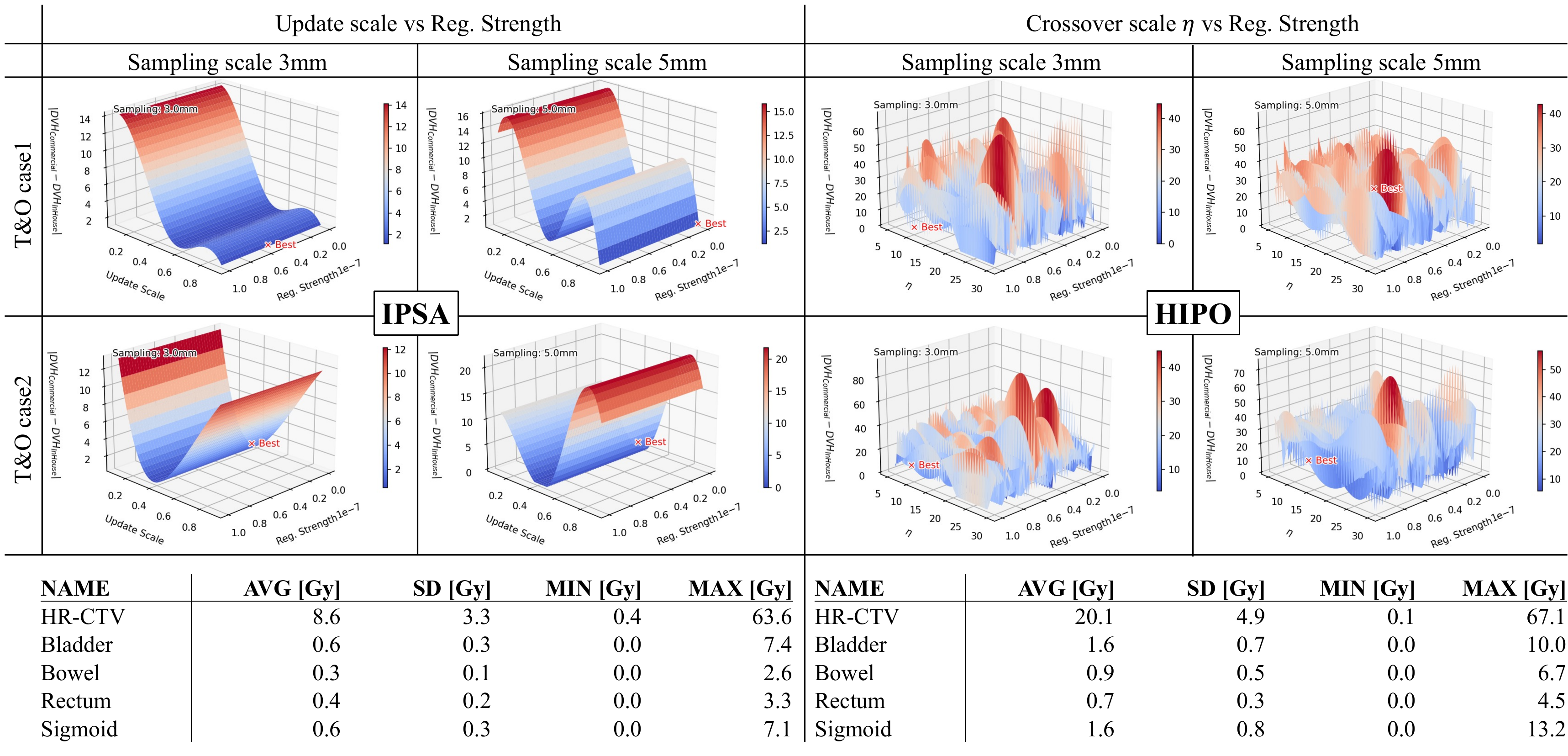}
    \caption{Absolute cumulative DVH differences for 18 T\&O cases across variations of three hidden parameters. Two T\&O cases are selected to visualize representative patterns. `Best' denotes the best case within the sampled hidden parameter combinations.}
    \vspace{-0.5cm}
    \label{fig:parameter_tuning_TO}
\end{figure}

\begin{figure}[t]
    \includegraphics[width=\columnwidth]{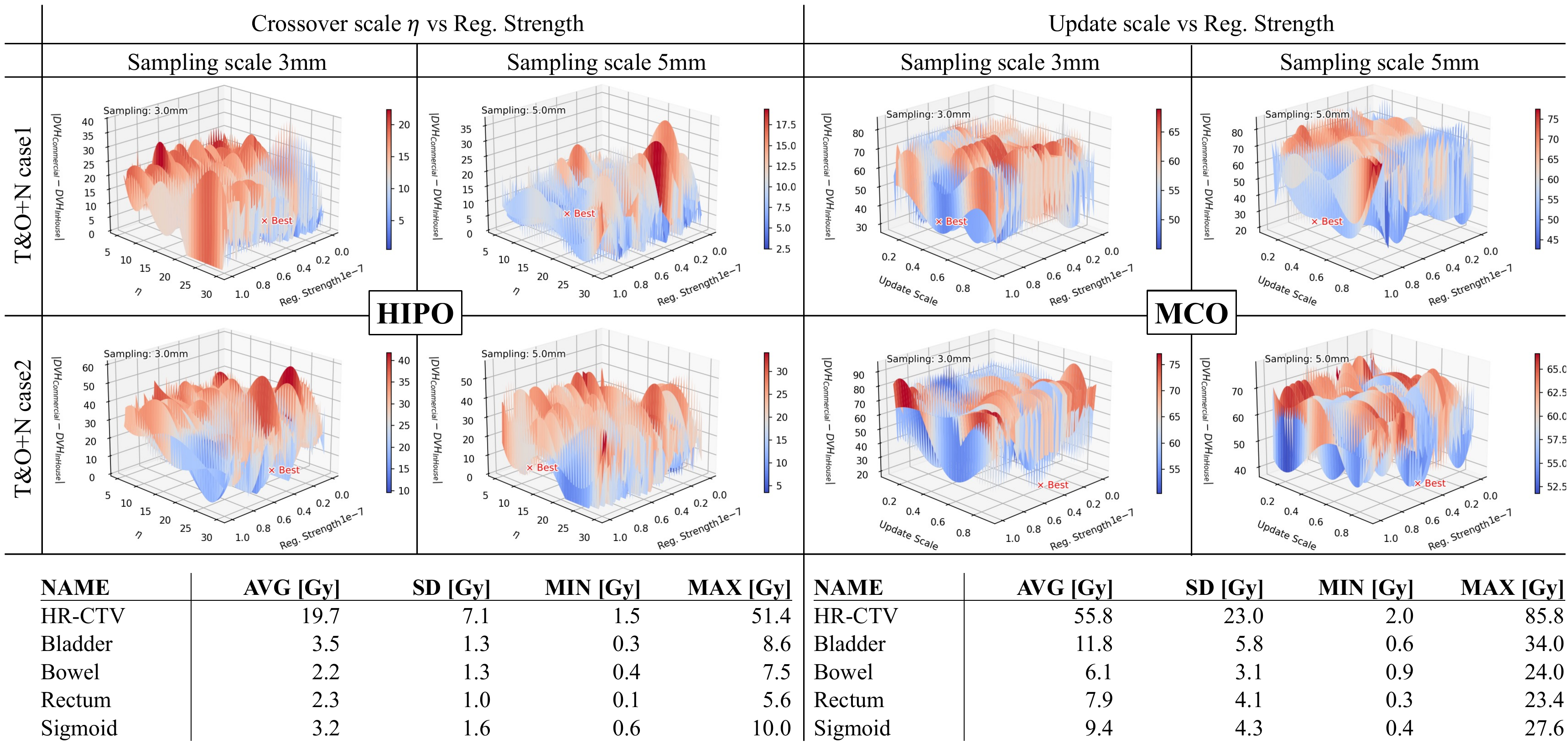}
    \vspace{-0.5cm}
    \caption{Absolute cumulative DVH differences for 6 T\&O+N cases across variations of three hidden parameters.Two T\&O+N cases are selected to visualize representative patterns.}
    \label{fig:parameter_tuning_TON}
\end{figure}

\begin{table}[t]
    \caption{DVH matching rates between commercial (baseline) and in-house inverse optimization algorithms using a 5\% volume-difference tolerance.}
    \vspace{0.15cm}
    \label{tab:dvh_matching}
    \resizebox{\columnwidth}{!}{%
    \begin{tabular}{l|l| ccccc}
        \hline
            Applicator & IO & HR-CTV & Bladder & Bowel & Rectum & Sigmoid \\
        \hline
            T\&O   & IPSA & 78.3\% & 99.3\% & 99.1\% & 99.2\% & 98.9\% \\
            T\&O+N & HIPO & 49.2\% & 92.4\% & 95.1\% & 95.9\% & 92.4\% \\
        \hline
    \end{tabular}
    }
\end{table}

\section{Methods}
\subsection{Clinical cases and applicator types}
A total of 24 cervical cancer cases previously treated with HDR brachytherapy were retrospectively analyzed. The cohort consisted of 18 tandem-and-ovoid (T\&O) cases and 6 tandem-and-ovoid with needle (T\&O+N) cases. Contours of high-risk clinical target volume (HR-CTV) and OAR structures (bladder, rectum, sigmoid, and bowel) were available for all cases.

\begin{table*}[t]
    \caption{Case-by-case absolute DVH metric differences (Gy) for T\&O (IPSA vs. HIPO) and T\&O+N (HIPO vs. MCO) cases.}
    \vspace{0.15cm}
    \label{tab:casewise_dvh}

    \centering
    \footnotesize
    \renewcommand{\arraystretch}{0.7}
    \setlength{\tabcolsep}{4pt}
    % \resizebox{\textwidth}{!}{%
    \begin{tabular}{c|c | rrrrrr | rrrrrr}
        \toprule
        \multirow{2}{*}{\textbf{Applicator}} & \multirow{2}{*}{\textbf{Case}} & \multicolumn{6}{c|}{\textbf{IPSA (for T\&O) / HIPO (for T\&O+N)}} & \multicolumn{6}{c}{\textbf{HIPO (for T\&O) / MCO (for T\&O+N)}} \\
        & & \textbf{D98} & \textbf{D90} & \textbf{Bladder} & \textbf{Bowel} & \textbf{Rectum} & \textbf{Sigmoid} & \textbf{D98} & \textbf{D90} & \textbf{Bladder} & \textbf{Bowel} & \textbf{Rectum} & \textbf{Sigmoid} \\
        \midrule
        \multirow{18}{*}{T\&O} 
        & 1  & 0.2 & 0.6 & 0.4 & 1.4 & 0.1 & 0.9 & 2.0 & 2.8 & 1.7 & 2.0 & 0.3 & 1.6 \\
        & 2  & 0.0 & 0.8 & 0.1 & 0.9 & 0.0 & 0.0 & 3.0 & 3.0 & 0.9 & 0.4 & 0.2 & 0.5 \\
        & 3  & 0.5 & 0.3 & 0.6 & 0.2 & 0.3 & 0.0 & 1.5 & 0.4 & 1.2 & 0.4 & 0.6 & 0.1 \\
        & 4  & 0.7 & 0.7 & 0.3 & 0.2 & 0.1 & 0.2 & 5.5 & 7.4 & 1.4 & 3.4 & 0.3 & 1.1 \\
        & 5  & 0.6 & 0.1 & 0.2 & 0.0 & 0.2 & 0.1 & 2.0 & 3.6 & 2.7 & 0.4 & 0.6 & 2.1 \\
        & 6  & 5.5 & 7.1 & 2.9 & 0.0 & 1.6 & 3.9 & 5.2 & 6.6 & 2.8 & 0.0 & 1.6 & 3.7 \\
        & 7  & 0.0 & 0.3 & 0.0 & 0.3 & 0.0 & 0.2 & 6.9 & 9.1 & 5.7 & 3.5 & 0.6 & 2.7 \\
        & 8  & 0.9 & 0.4 & 0.3 & 0.4 & 0.4 & 0.0 & 2.8 & 4.3 & 2.8 & 1.5 & 0.5 & 2.1 \\
        & 9  & 1.1 & 0.5 & 0.1 & 0.6 & 0.2 & 0.0 & 4.9 & 6.6 & 4.0 & 1.9 & 0.7 & 1.9 \\
        & 10 & 0.2 & 0.1 & 0.7 & 1.0 & 0.2 & 0.1 & 8.5 & 10.8 & 8.7 & 5.8 & 0.7 & 3.5 \\
        & 11 & 1.3 & 2.6 & 1.6 & 0.3 & 0.1 & 0.6 & 4.2 & 6.2 & 4.1 & 0.9 & 0.9 & 1.9 \\
        & 12 & 0.8 & 0.2 & 0.2 & 0.1 & 0.2 & 0.5 & 3.3 & 5.6 & 2.7 & 0.6 & 0.4 & 2.7 \\
        & 13 & 1.8 & 1.6 & 0.2 & 0.3 & 0.4 & 0.2 & 0.5 & 1.6 & 1.4 & 2.1 & 0.3 & 1.6 \\
        & 14 & 0.4 & 0.1 & 0.1 & 0.1 & 0.0 & 0.8 & 2.2 & 3.7 & 1.4 & 0.4 & 0.9 & 2.1 \\
        & 15 & 0.7 & 0.1 & 0.1 & 1.1 & 0.2 & 0.2 & 0.7 & 0.3 & 0.4 & 0.3 & 0.3 & 0.1 \\
        & 16 & 1.4 & 1.7 & 0.7 & 1.5 & 0.6 & 0.1 & 3.2 & 5.5 & 10.2 & 8.1 & 2.3 & 3.0 \\
        & 17 & 0.4 & 0.3 & 0.3 & 0.1 & 0.3 & 0.1 & 0.0 & 0.1 & 0.2 & 0.0 & 0.0 & 0.0 \\
        & 18 & 2.2 & 2.4 & 0.8 & 0.8 & 0.5 & 0.4 & 1.4 & 1.0 & 0.5 & 0.2 & 0.3 & 0.2 \\
        \midrule
        \multirow{6}{*}{T\&O+N} 
        & 1  & 9.1 & 10.7 & 7.3 & 4.6 & 3.4 & 4.9 & 20.3 & 23.9 & 12.9 & 6.5 & 4.5 & 7.7 \\
        & 2  & 6.3 & 8.8  & 5.0 & 2.0 & 2.4 & 4.3 & 14.9 & 17.6 & 9.5  & 2.3 & 3.4 & 6.3 \\
        & 3  & 2.6 & 3.2  & 2.6 & 2.0 & 1.9 & 2.1 & 28.4 & 33.3 & 17.0 & 6.6 & 8.6 & 14.8 \\
        & 4  & 7.6 & 9.0  & 9.6 & 3.6 & 4.7 & 6.0 & 13.3 & 15.8 & 15.4 & 3.8 & 7.1 & 8.8 \\
        & 5  & 3.9 & 4.1  & 1.5 & 0.0 & 0.3 & 1.3 & 2.3  & 2.8  & 1.7  & 0.0 & 0.5 & 1.1 \\
        & 6  & 14.3 & 17.4 & 5.4 & 7.0 & 3.5 & 9.5 & 13.6 & 17.1 & 3.9  & 6.6 & 2.6 & 7.4 \\
        \bottomrule
    \end{tabular}%
    % }
\end{table*}

\subsection{Commercial TPS planning}
Commercial plans were generated using Oncentra Brachy (v4.6.3) with user-defined input parameters. Input parameters (dose objectives and relative weights) were identical across all commercial TPS plan generations and are summarized in Table~\ref{tab:input_params}. The resulting plans from commercial TPS served as reference plans for retrospective comparison.

\subsection{In-house IO with hidden parameters}
Three IO algorithms\textemdash{}IPSA, HIPO, and MCO\textemdash{}were implemented in-house. These implementations enabled explicit control of hidden input parameter settings that are not exposed in commercial TPS implementations.

User-defined input parameters were fixed to match those used in the commercial TPS. Internal algorithm-specific parameters, including dwell-time modulation constraints, convergence thresholds, optimization tolerances, and penalty weights, were systematically varied. For IPSA and MCO, an additional dwell-time update scaling parameter was included. For HIPO, the crossover coefficient $\eta$ was included as an algorithm-specific parameter. Each IO algorithm was evaluated using 1,000 combinations of hidden parameter settings.

Plan evaluation was performed using cumulative dose-volume histogram (DVH) curves and standard dosimetric indices, including HR-CTV D98/D90 and OAR D2cc. DVH matching rates were computed using a 5\% volume-difference tolerance. The dosimetric sensitivity to hidden parameter variation was quantified using absolute DVH differences and the standard deviation (SD) across parameter combinations.

\section{Results}
Figure~\ref{fig:plan_results} shows that cumulative DVH curves of commercial and in-house plans closely align with the clinical plan, yet spatial dose distributions diverge. These differences are not easily captured by DVH metrics. Table~\ref{tab:dvh_matching} summarizes DVH matching rates using a 5\% volume-difference tolerance. OAR agreement was high (92.0\%–99.3\%), but HR-CTV matching was lower—78.3\% for IPSA and only 49.2\% for HIPO—suggesting increased sensitivity of target coverage to hidden parameter variation. 

Figures~\ref{fig:parameter_tuning_TO} and~\ref{fig:parameter_tuning_TON} present absolute cumulative DVH differences across variations of three hidden parameters for 18 T\&O and 6 T\&O+N cases. Two surface plots per case illustrate the nonlinear response of DVH outcomes across the sampled parameter space. T\&O+N cases exhibited larger SDs and broader surface topography, confirming that applicator complexity heightens optimizer sensitivity. Table~\ref{tab:casewise_dvh} reports absolute DVH metric differences across all cases. The lowest variability in T\&O plans was observed with HIPO, while MCO showed the greatest spread in T\&O+N cases—but also included some of the most optimal solutions. Algorithms exhibiting larger SDs also showed a wider range of achievable DVH outcomes across the sampled parameter combinations.

\section{Discussion}
Most studies comparing inverse optimization (IO) algorithms have focused on plan quality under user-defined input parameters. In clinical TPS implementations, however, these algorithms are effectively treated as black boxes, with internal parameters embedded and not exposed to users.

By implementing IPSA, HIPO, and MCO in-house, this study enabled direct control over key internal settings, including dwell-time modulation constraints, convergence thresholds, optimization tolerances, and penalty weights. The results demonstrate that variations in these hidden input parameters can lead to substantial dosimetric differences, even when identical user-defined input parameters are applied, with HR-CTV D90 absolute differences reaching up to 28.4 Gy.

Algorithms exhibiting wider output distributions, characterized by higher standard deviations, were observed to include solutions with improved target coverage under specific parameter configurations. This indicates that such algorithms can yield improved plans when parameters are appropriately tuned. Without such tuning, however, plan consistency may degrade.

\section{Conclusions}
Hidden input parameter settings exert a substantial influence on dosimetric outcomes in GYN HDR brachytherapy across inverse optimization algorithms.
Even under identical user-defined dose objectives, systematic variation of internal parameters resulted in large differences in target coverage and organ-at-risk doses, with effects amplified in more complex applicator configurations.

By implementing IPSA, HIPO, and MCO in-house and evaluating their sensitivity to hidden parameter variation, this study provides a quantitative assessment of internal algorithm behavior that is not accessible in commercial TPS workflows.
These findings highlight that plan quality comparisons based solely on user-defined inputs may be insufficient and underscore the importance of internal parameter configurations when evaluating and applying IO algorithms for clinical HDR brachytherapy planning.

\balance

\end{document}